\documentclass[conference,onecolumn]{IEEEtran}
\usepackage{graphicx}
\usepackage{comment}
\usepackage{amssymb}

\ifCLASSINFOpdf
\else
\fi
\begin{document}
%
\title{Several Consequences of Optimality}

\author{
\IEEEauthorblockN{Dibakar~Das}
\IEEEauthorblockA{IIIT Bangalore\\
Email: dibakard@acm.org}
}


%


\maketitle

\begin{abstract}
Rationality is frequently associated with making the best possible decisions. It's widely acknowledged that humans, as rational beings, have limitations in their decision-making capabilities. Nevertheless, recent advancements in fields, such as, computing, science and technology, combined with the availability of vast amounts of data, have sparked optimism that these developments could potentially expand the boundaries of human bounded rationality through the augmentation of machine intelligence. In this paper, findings from a computational model demonstrated that when an increasing number of agents independently strive to achieve global optimality, facilitated by improved computing power, etc., they indirectly accelerated the occurrence of the "tragedy of the commons" by depleting shared resources at a faster rate.  Further, as agents achieve optimality, there is a drop in information entropy among the solutions of the agents. Also, clear economic divide emerges among agents. Considering, two groups, one as producer and the other (the group agents searching for optimality) as consumer of the highest consumed resource, the consumers seem to gain more than the producers. Thus, bounded rationality could be seen as boon to sustainability.
\end{abstract}


%
\IEEEpeerreviewmaketitle

\section{Introduction}



Rationality, particularly the idea of perfect rationality, is commonly associated with agents consistently making optimal decisions. The concept of rationality has been a subject of extensive debate for many years. Its definition varies across various disciplines, including cognition, philosophy, psychology, economics, as well as mathematical and computational theories \cite{cite_scope_of_rationality}. Several theories, such as, game theory, were initially built on the assumption of perfect rationality among participants. In computational and mathematical contexts, perfect rationality typically entails agents (such as, humans) always maximizing their expected utility to achieve the global optimum in a given problem \cite{cite_rationality_as_maximize_utility}. However, the concept of perfect rationality has been challenged by the idea of bounded rationality \cite{cite_handbook_of_bounded_rationality}. In bounded rationality, human decision-making is characterized by not always reaching globally optimal solutions due to inherent biases and limited computational resources. Mathematically, this is expressed as suboptimal decisions, or local optima.

With the recent progress in science and technology, including high-performance computing, artificial intelligence and data science, there is a growing interest in research focused on expanding the boundaries of bounded rationality to enable humans to achieve higher levels of rational decision-making \cite{cite_ai_will_enhance_the_bounds_bounded_rationality_marwala}. The emerging field of \emph{computational rationality} \cite{cite_computational_rationality_review} represents a significant step in this direction. Computational rationality is fostering the integration of disciplines, such as, cognition, artificial intelligence and computational neuroscience, with the goal of enabling agents to consistently maximize their expected utility. These advancements across various domains imply that humans will have the capacity to more rapidly and consistently attain global optimum, ultimately achieving this with greater regularity through the augmentation of machine intelligence.


Sustainability consists of three key facets, namely, environment, economy and society \cite{cite_sustain_env_economy_society}. Global sustainable development has been a central focus for several decades, primarily driven by concerns regarding the precarious state of the Earth's environment \cite{cite_unesco_sustainable_development}. Effective and efficient resource management is vital for achieving sustainability. The emerging field of \emph{computational sustainability} revolves around the use of computational and mathematical techniques to optimize the utilization of environmental, economic, and social resources \cite{cite_unesco_sustainable_development}. Unrestricted exploitation of resources by individuals or groups can lead to the depletion of these resources, a phenomenon often described as the "tragedy of the commons" \cite{cite_tragedy_of_the_commons_hardin}.


This article aims to draw attention to a pressing issue of whether the increased speed at which a growing number of agents, equipped with the advancements mentioned in various fields, can attain global optimal solutions to the same problem might actually accelerate the occurrence of the tragedy of the commons. To investigate this, a basic computational model composed of a cluster of agents attempting to independently solve the same problem (specifically, the Knapsack problem) using genetic algorithms (GA) is employed. The results indicate that as these agents achieve global optimality more rapidly, thanks to enhanced computing power and data resources, it paradoxically leads to a swifter emergence of the tragedy of the commons. Further, as agents progress towards global optimality, there is a decrease in the information entropy of solutions among them, and a distinct economic disparity is observed. When considering two groups, one acting as producers and the other (comprising agents in pursuit of optimality) as consumers of the most heavily utilized resource, it appears that the consumers benefit more than the producers in this scenario.

\section{The problem with an example}
Let's break down the problem using a simple example. Imagine a group of agents independently making pizzas and they have a set of ingredients $\{a, b, ..., z\}$ to choose from. The most delicious pizza is created when they use the ingredients $\{a, e, i, o, u\}$ but let's assume the agents are unaware of this ideal ingredient combination. Each agent tries to make pizzas based on their local conditions, knowledge and the availability of ingredients (referred to as local optima). Consequently, some agents may reach the global optimal set $\{a, e, i, o, u\}$ while many others won't. This is because there are a total of $2^{26}$ (exponential) possibilities when using a single bit to represent whether an item from the set $\{a, b, ..., z\}$ is chosen or not.

In this scenario, the utilization of the resources $\{a, e, i, o, u\}$ is not heavily stressed since only a few agents would reach this combination. However, if these agents were equipped with enhanced capabilities, such as, improved computing power, machine intelligence, advanced algorithms and other innovations from the various fields mentioned earlier, most of them would quickly converge on the global optimal solution $\{a, e, i, o, u\}$. As a result, more and more agents would use these resources leading to overconsumption.

Eventually, when all agents adopt this global optimal solution they would exhaust the ingredients in the set $\{a, e, i, o, u\}$ which would lead to tragedy of the commons. In essence, the accelerated attainment of global optimality aided by advancements in fields like computing and machine intelligence would hasten the tragedy of the commons.

Interestingly, bounded rationality which involves agents not always pursuing the global optimal solution turns out to be a hidden advantage for sustainability. This is because it introduces diversity in the solutions, albeit suboptimal ones, rather than forcing all agents onto the same global optimal solution.

It's worth noting that this paper expands upon a scenario in which an increase in resource consumption occurs due to heightened demand resulting from reduced resource costs and from enhanced production efficiency \cite{cite_jevons_paradox_5}. What's particularly significant is this paper's focus on how the process of multiple agents independently striving to reach global optimality for the same problem more quickly with improved computing capabilities can accelerate the depletion of resources. This phenomenon can occur independent of the cost of the resource or production efficiency. The paper also highlights how the diversity of solutions represented by the local optima of different agents plays a crucial role in mitigating this issue. As will be seen latter in the results that there is a decline in the information entropy of solutions among the agents and a clear economic divide emerges. If two groups are considered, one as producers and the other (comprising agents trying to reach optimality) as consumers of the most heavily utilized resource, it appears that the consumers benefit more than the producers in this scenario.



\section{Computational Model}\label{section_computation_model}
The computational model involves a group of agents, each attempting to independently solve the same NP-complete problem. Specifically, they are tackling the Knapsack problem, which falls into the category of NP-complete problems \cite{cite_knapsack_6}. The primary goal of the Knapsack problem is to choose a selection of items, each with an assigned value and weight, to be placed into a sack with a defined capacity. The aim is to maximize the total value of the items within the constraints of the sack's capacity. In this paper, the scenario considers the complete inclusion of items in the sack which is known as the 0/1-Knapsack problem.


Consider a scenario where there are $N$ agents, and each of them is independently attempting to solve the same 0/1-Knapsack problem. This particular Knapsack problem involves $M$ items, characterized by a weight vector $[w_1, w_2, ..., w_M]$, a value vector $[v_1, v_2, ..., v_M]$ and a knapsack with a maximum weight capacity of $W$. The 0/1-Knapsack problem is defined as

\begin{equation}\label{eqn_knapsack_objective}
maximize \sum_{i=1}^M x_iv_i
\end{equation}
subject to
\begin{equation}\label{eqn_knapsack_constraint}
\sum_{i=1}^M x_iw_i \le W
\end{equation}
where $x_i \in \{0,1\}$.
In essence, the solution to this problem comes down to selecting a binary vector $X^{(opt)}$ of $x_i$s comprising 0s and 1s from a total of $2^M$ possible combinations. This selected vector should aim to maximize the objective function (\ref{eqn_knapsack_objective}) while adhering to the constraint (\ref{eqn_knapsack_constraint}).

The Knapsack problem is a thoroughly investigated problem that can be effectively solved using dynamic programming (DP) \cite{cite_knapsack_6}. Additionally, there are alternative approaches, such as, genetic algorithms (GA) have been proposed to tackle this problem  \cite{cite_ga_knapsack_7}. In this context, it is assumed that the agents do not have the knowledge dynamic programming (DP) or lack the capability to do so. Instead, they employ a trial-and-error method to solve the problem. This trial-and-error approach is implemented using a genetic algorithm (GA) because it simulates an evolutionary process where agents gradually improve their optimality over successive iterations or generations in GA.


Each agent, denoted with $j$, initiates its genetic algorithm (GA) with an initial random population consisting of vectors of 1s and 0s, each of length $M$, serving as the starting values for the solution vector $X^{(j)}$. The size of this initial population is set to $N_p^{(j)}$. From this population, $N_s^{(j)}$ vectors that meet the constraints are selected as potential solution candidates. Among these $N_s^{(j)}$ candidates, there may exist a superior local optimum or even the global optimum for agent $j$. If the size of $N_s^{(j)}$ is less than 2, a new initial population $N_p^{(j)}$ is generated.

In each generation $k$, agent $j$ then randomly selects two vectors from $N_s^{(j)}$, let's call them $X^{(j)(k)}_m$ and $X^{(j)(k)}_n$ and applies crossover. In the crossover process, the upper half of $X^{(j)(k)}_m$ is combined with the lower half of $X^{(j)(k)}_n$ to create a new vector $X^{(j)(k)}_p$. Some bits of $X^{(j)(k)}_p$ are subjected to mutation where specific bits are toggled.

These steps of random selection, crossover and mutation are repeated multiple times to form a new population $N_p^{(j)}$ which will be utilized in the next generation. This entire process is iterated over $N_g^{(j)}$ generations by each agent $j$ until it reaches a satisfactory local optimum solution or attains the global optimum $X^{(opt)}$. In each generation $k$, an agent will either achieve the global optimum or settle for a local optimum that complies with the given constraints. Over the course of multiple generations, each agent strives to enhance its current solution by finding a superior local optimum or reaching the global optimum.


If item $i$ is in the solution then $w_i$ is the amount of corresponding resource consumed. The cumulative resources consumed over all the generations is given by,
\begin{equation}
\sum_{i=1}^M\sum_{j=1}^N\sum_{k=1}^{N_g^{(j)}}x_i^{(j)(k)}w_i
\end{equation}
where $x_i^{(j)(k)} \in \{0,1\}$.

If the combined use of resource $i$ exceeds the total amount of that resource available, all agents relying on this resource whether it's part of their local or global optimal solutions will face a shortage of that resource. This situation, known as the tragedy of the commons, is accelerated when agents achieve global optimal solutions more quickly due to improved computational abilities. In other words, the faster agents reach these global optimal solutions the sooner they deplete resources, intensifying the tragedy of the commons. This underscores the crucial need for sustainable resource management in a world where computational capabilities are advancing and resource demand is on the rise.
\subsection{Information entropy of the system}
Information entropy defines the amount of randomness in the system. The more the randomness the more the diversity in the system. It is interesting to find out how the (information) entropy of agents in the system changes with optimality. Solutions of all agents in a generation are represented as a contingency table and then the entropy is evaluated. It will be observed latter that with the progress towards optimality the entropy starts decreasing. Thus, drop in information entropy could lead to tragedy of the commons as an externality.
\subsection{Utility of achieving optimality}
Each agent has a certain benefit on achieving optimality. Let $U^{(opt)}$ be the utility of an agent each time it uses the global optimal solution during $N_g$ generations (assuming $N^{(j)}_g = N_g$ for all agents). The earlier an agent achieves the global optimal solution the higher is its total utility. It would be interesting to see how the utility of agents behaves over time (i.e., generations). Let $N^{(opt)}_j$ be the number of times agent $j$ uses the optimal solution. Then, utility of $j^{th}$ agent, $u^{(opt)}_j$ is,
\begin{equation}
u^{(opt)}_j = N^{(opt)}_j U^{(opt)}
\end{equation}
As will be observed latter, optimality could lead to economic divide. If the utility is exponential in time (i.e. $N_g$), it can lead to dominance by a few agents with windfall gains for early achievers. Then, the utility of agent is,
\begin{equation}
u^{{(opt)}_e}_j = \sum^{\|S^{(j)}_g\|}_{l \in \{S^{(j)}_g\}} U^{(opt)}e^{\frac{N_g - l}{a}}
\end{equation}
where $S^{(j)}_g$ is the set of generations where agent $j$ exercised the optimal solution, ${\|S^{(j)}_g\|}$ is the cardinality of $S^{(j)}_g$ and $a$ is a control parameter.
\subsection{Producer and consumer utilities}
This section models the utilities of the highest consumed resource for two groups of agents, namely, the consumers and the producers of the concerned resource. The first group is the original set of $N$ agents who derive a certain utility by using the highest consumed resource. The second  is a group of $P$ agents of producers who gain by producing the same highest consumed resource.
Let $i_{max}$ be index of the highest consumed resource. Thus, $w_{i_{max}}$ is the amount of highest resource that may be consumed by an agent in each generation. In generation $k$, the resource consumed is given by,
\begin{equation}
R_{i_{max}}^{(k)} = w_{i_{max}} \sum_{j=1}^{N}x^{(j)(k)}_{i_{max}}
\end{equation}
where $x^{(j)(k)}_{i_{max}} \in \{0,1\}$ is value in the solution vector $X^{(j)(k)}$ of agent $j$ at generation $k$ indexed by ${i_{max}}$. Thus, the utility of the consumers in generation $k$ is given by,
\begin{equation}
U^{(c)} = \alpha R_{i_{max}}^{(k)}
\end{equation}
where $\alpha, 0 \le \alpha \le 1$ is the utility factor.

For the producers, the utility is based on two things. Firstly, the value they get by selling the amount consumed by the consumers and secondly, the reserve of remaining resource. The utility of the producers is given by,
\begin{equation}
U^{(p)} = \beta R_{i_{max}}^{(k)} + \gamma^{k+1} \beta \left(T_{R_{i_{max}}} - \sum_{m=1}^{k-1} R_{i_{max}}^{(m)}\right)
\end{equation}
where $\beta, 0 \le \beta \le 1$ is the utility factor, $\gamma, 0 \le \gamma \le 1$ is the discount factor (discounting uncertain future gains) and $T_{R_{i_{max}}}$ is the total reserve of the maximum consumed resource.
\section{Results}\label{section_results}
In this section, we present the simulation results using the aforementioned computational model. The parameters utilized in the simulation are detailed in Table \ref{table_simulation_parameters}. The simulation was conducted using the \verb|Python| programming language.

Firstly, the three different scenarios of resource consumption are discussed:
\begin{enumerate}
\item \emph{All agents striving for global optimum}: In this scenario, every agent is attempting to achieve the global optimum.

\item \emph{Most agents content with local optimum}: Here, a significant portion of agents is content with achieving local optima.

\item \emph{Some agents achieve faster global optimum}: In this scenario, certain agents possess the ability to reach the global optimum more rapidly due to their enhanced computing capabilities.
\end{enumerate}

\begin{table}[ht]\label{table_simulation_parameters}
  \caption{Simulation Parameters}
  \centering
  \begin{tabular}{|p{2cm}|p{2cm}|p{3cm}|}
  \hline
  Parameter & Value & Description\\  [0.5ex]
  \hline
  $N$ & 25 & Number of agents\\
  \hline
  $M$ & 10 & Number of items\\
  \hline
  $w_1, w_2,..., w_{10}$ & 996, 771, 543, 593, 621, 473, 595, 388, 935, 874 & Weights of items, uniform randomly generated\\
  \hline
  $v_1, v_2,..., v_{10}$ & 54.04769411, 39.33601431, 14.83657681, 43.52375770, 66.31920392, 26.17907976, 27.14489409, 58.72956010, 25.50253249, 49.04678721 & Value of items, uniform randomly generated\\
  \hline
  $W$  & $0.5 \times \sum_{i=1}^{10}w_i$ & Capacity of sack\\
  \hline
  $X^{(opt)}$ & 1,1,0,1,1,0,0,1,0,0 & Optimal solution using dynamic programming for validation only\\
  \hline
  $X^{(opt)}$ & 1,1,0,1,1,0,0,1,0,0 & Optimal solution using dynamic programming for validation only\\
  \hline
  $N_p^{(j)}, j = 1,2,..,25$  & 45 & Initial population size. Assumed same for all agents\\
  \hline
  $N_s^{(j)}, j = 1,2,.., 25$  & 45 & Probable solution set size. Assume same for all agents\\
  \hline
  $N_g^{(j)}, j = 1,2,.., 25$  & 2000 & Number of generations each agent tries to achieve better or optimal solution. Assumed same for all agents\\
  \hline
  \end{tabular}
  \label{table_model_parameters}
\end{table}

\begin{figure}[ht]
\centering
\includegraphics[width=\columnwidth]{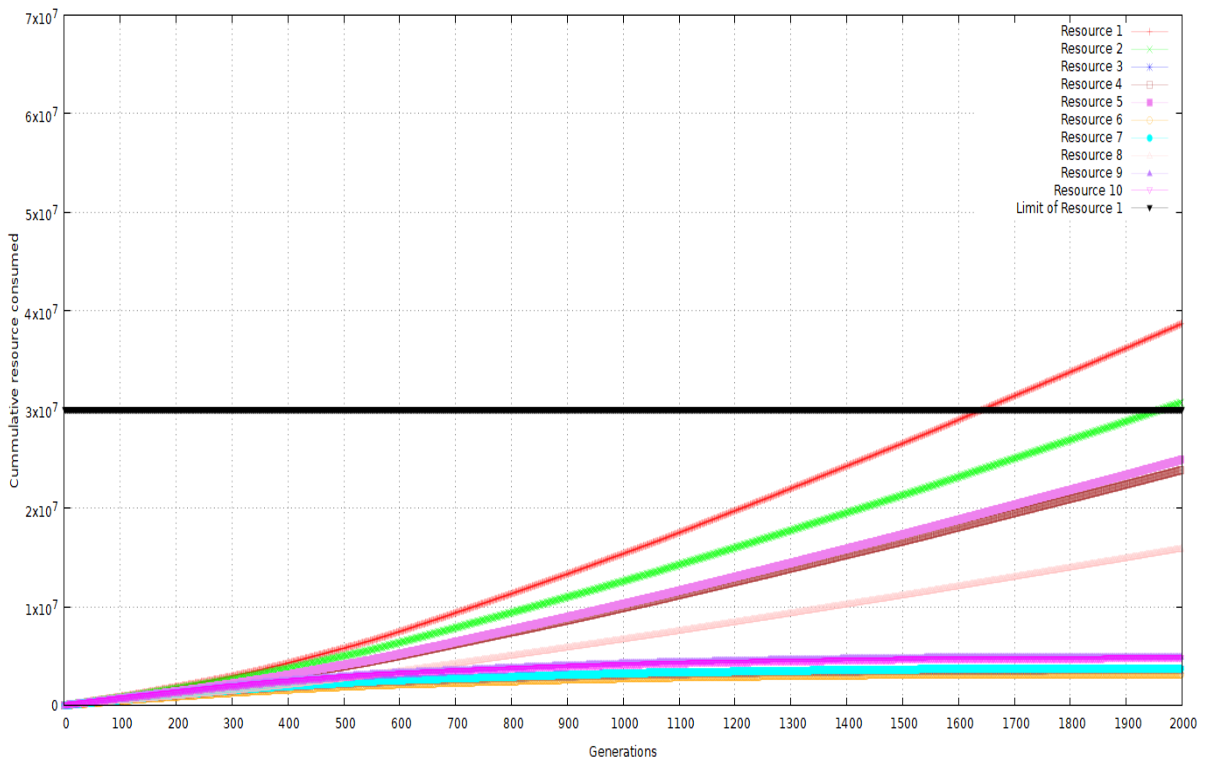}
\caption{Resource consumption (first scenario)}
\label{fig_total_resource_with_GA_inkscape}
\end{figure}

\begin{figure}[ht]
\centering
\includegraphics[width=\columnwidth]{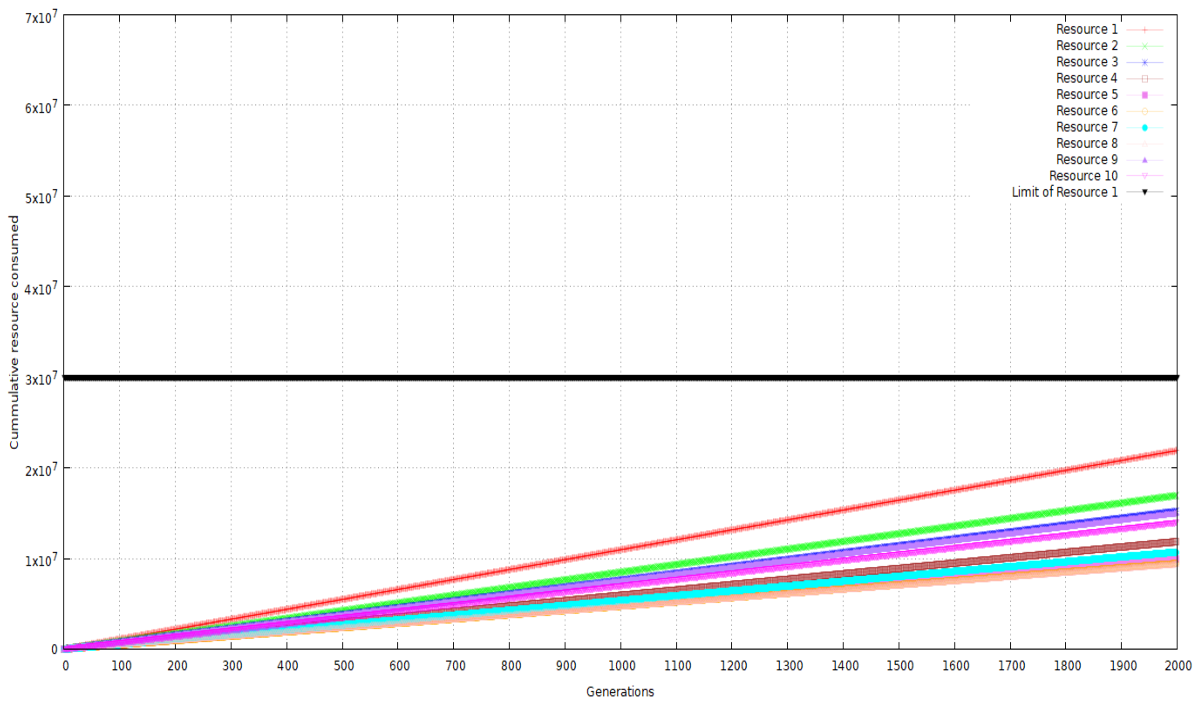}
\caption{Resource consumption (second scenario)}
\label{fig_total_resource_sub_opt_GA_inkscape}
\end{figure}

\begin{figure}[ht]
\centering
\includegraphics[width=\columnwidth]{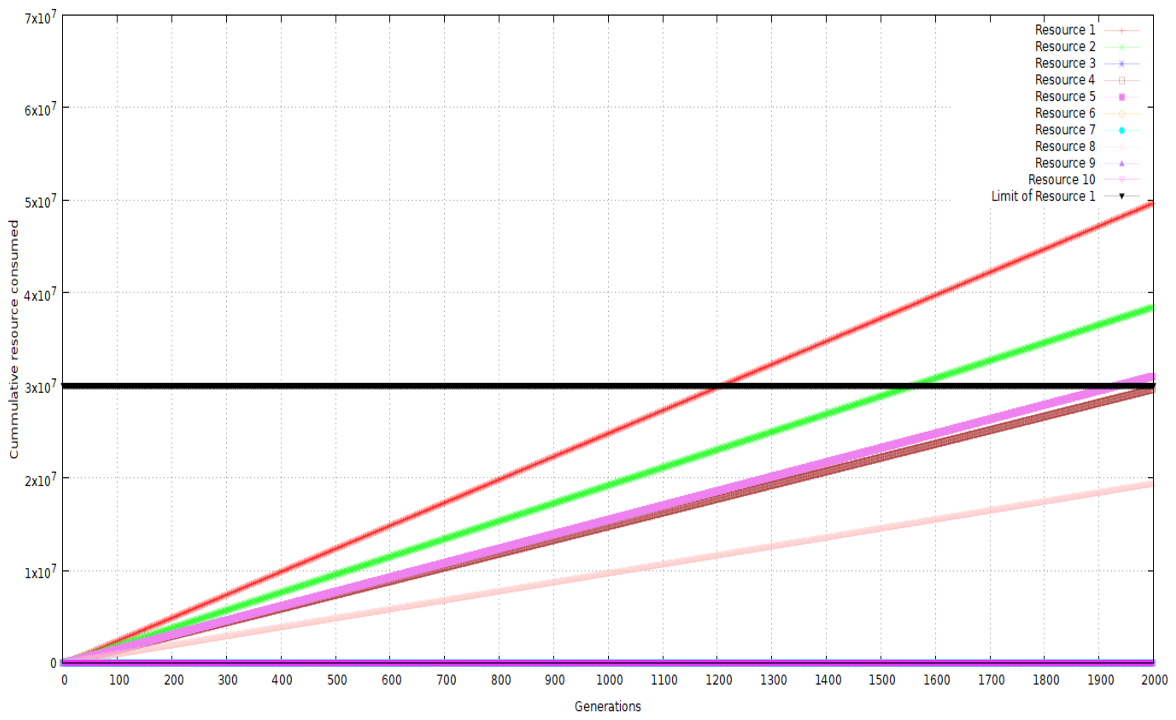}
\caption{Resource consumption (third scenario)}
\label{fig_total_resource_accelerate_GA_inkscape}
\end{figure}

\begin{figure}[ht]
\centering
\includegraphics[width=\columnwidth]{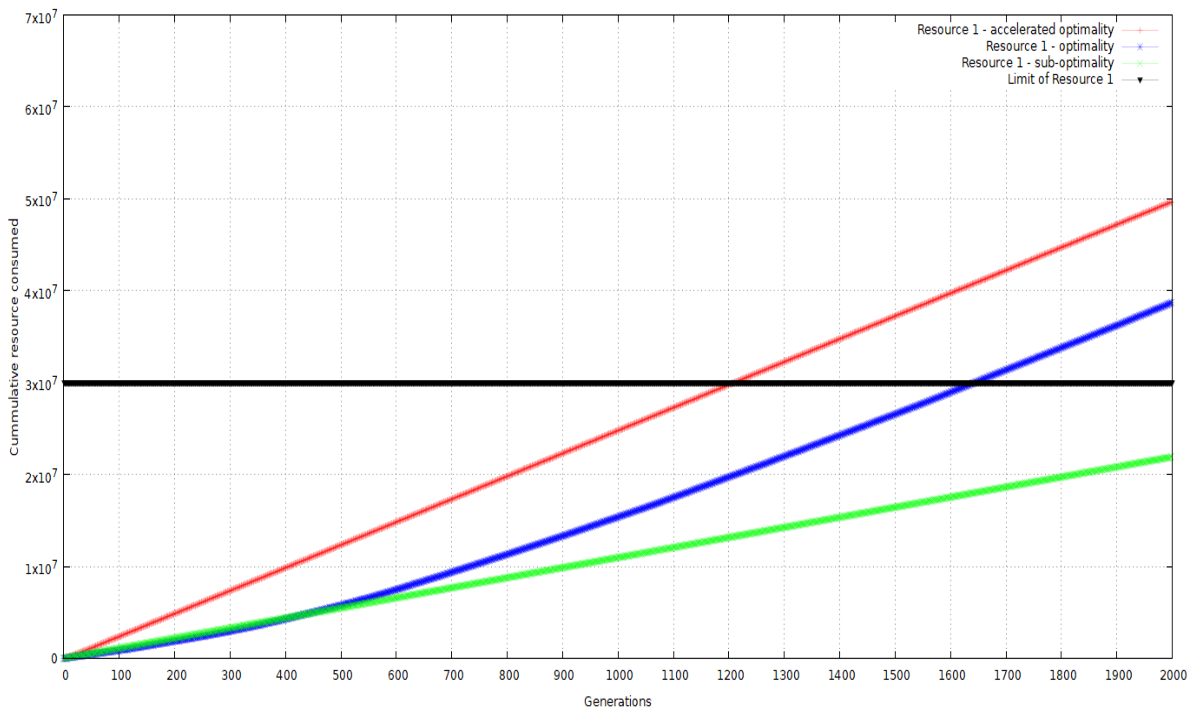}
\caption{Highest resource consumed, i.e., Resource 1 (all 3 scenarios)}
\label{fig_total_resource_compare_3_scenarios_inkscape}
\end{figure}

In the first scenario, the resource consumption is depicted in Fig. \ref{fig_total_resource_with_GA_inkscape}. The cumulative resources consumed are plotted along the \emph{y}-axis, and generations are represented along the \emph{x}-axis, consistent with all subsequent figures (Figs. \ref{fig_total_resource_with_GA_inkscape}-\ref{fig_total_resource_compare_3_scenarios_inkscape}). It's evident from the graph that the consumption of the most heavily used resource surpasses the resource availability threshold at approximately generation 1600. This outcome occurs when all agents successfully attain the global optimum.

In the second scenario, depicted in Fig. \ref{fig_total_resource_sub_opt_GA_inkscape}, a significant portion of agents are content with achieving local minima. It is notable that for the same number of generations the cumulative resource consumption by the agents remains well below the resource availability threshold. In this case, none of the agents manage to reach the global optimum.

In the third scenario, as illustrated in Fig. \ref{fig_total_resource_accelerate_GA_inkscape}, some random agents possess the capability to achieve the global optimum more rapidly due to their enhanced computing capabilities. It is evident that the consumption of the most heavily used resource surpasses the resource availability threshold at approximately generation 1200. This is notably much earlier compared to Fig. \ref{fig_total_resource_with_GA_inkscape} where all agents were striving for global optimum and reached resource availability threshold at generation 1600. In this case as well, all agents but one manage to achieve the global optimum much earlier than in the first scenario depicted in Fig. \ref{fig_total_resource_with_GA_inkscape}.

Fig. \ref{fig_total_resource_compare_3_scenarios_inkscape}  provides a clear representation of the cumulative highest consumed resource extracted from Figs. \ref{fig_total_resource_with_GA_inkscape}-\ref{fig_total_resource_accelerate_GA_inkscape} making it easier to draw conclusions. It is evident that as agents strive for the global optimum while attempting to achieve it more rapidly due to their enhanced computing capabilities accelerates resource consumption. This outcome leads to an expedited tragedy of the commons highlighting the relationship between the pace of resource consumption and the agents' pursuit of the (accelerated) global optimum for the same problem.

\subsection{Agents' optimality satisfaction}
\begin{figure}[ht]
\centering
\includegraphics[width=\columnwidth]{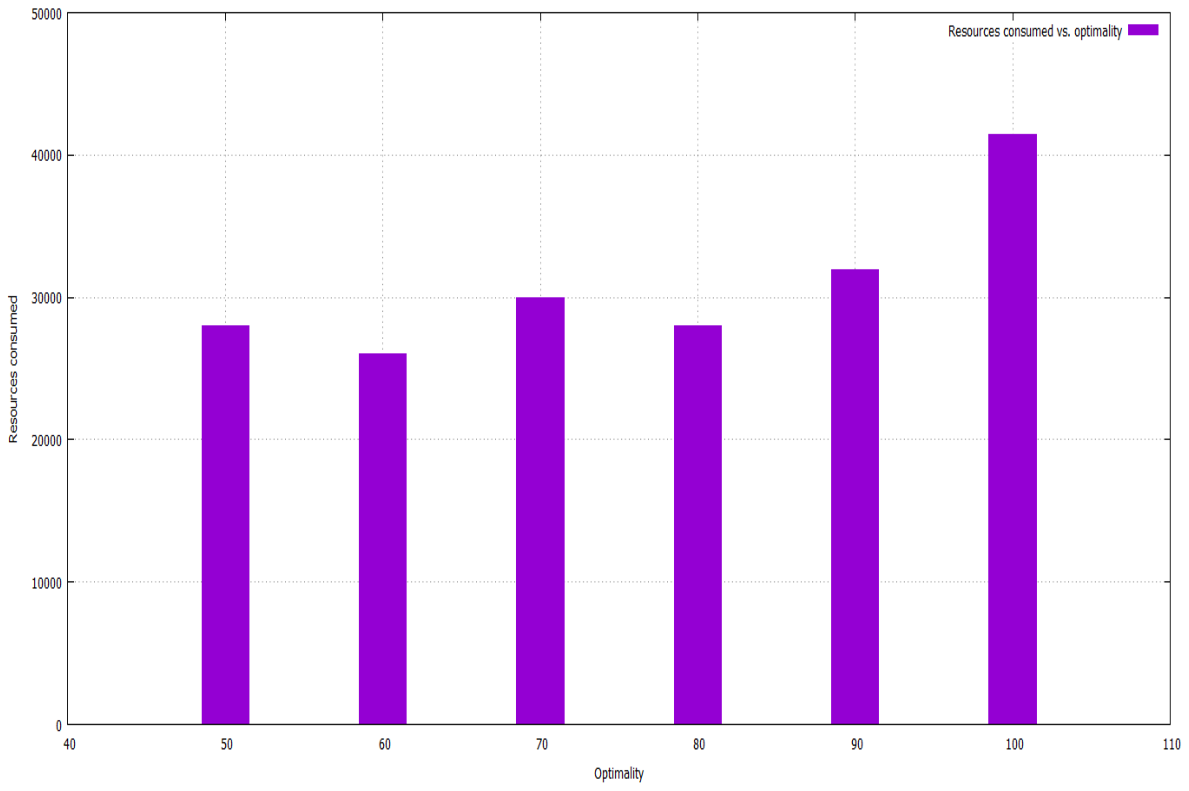}
\caption{Optimality satisfaction of agents}
\label{fig_satisfaction_inkscape}
\end{figure}

As observed above, agents operating in optimality can lead to higher resource consumption. If agents operate at lower optimality, it may lead to lesser resource consumption. Fig. \ref{fig_satisfaction_inkscape} shows resource consumption when agents are satisfied achieving $50\%$ to $100\%$ of their global optimal solutions. The satisfaction levels and resource consumptions are shown along \emph{x} and \emph{y}-axes respectively. Consumption of resources for $50\%$ to $80\%$ is $\leqslant 3 \times 10^4$. For 90\% satisfaction level, the consumption increases by $7\%$  from $3\times10^4$. However, achieving $100\%$ optimality leads to increase of 30\% resource consumption than the $90\%$ satisfaction level. Thus, if agents are less greedy and satisfy themselves at $10\%$ below the global optimality, this can lead to much lesser consumption of resources leading to prolonged sustainability. This situation gives greater chances for disruptive technologies and other policy interventions to play a role and avoid or at least delay tragedy of the commons.
\subsection{Information entropy behaviour}
\begin{figure}[ht]
\centering
\includegraphics[width=\columnwidth]{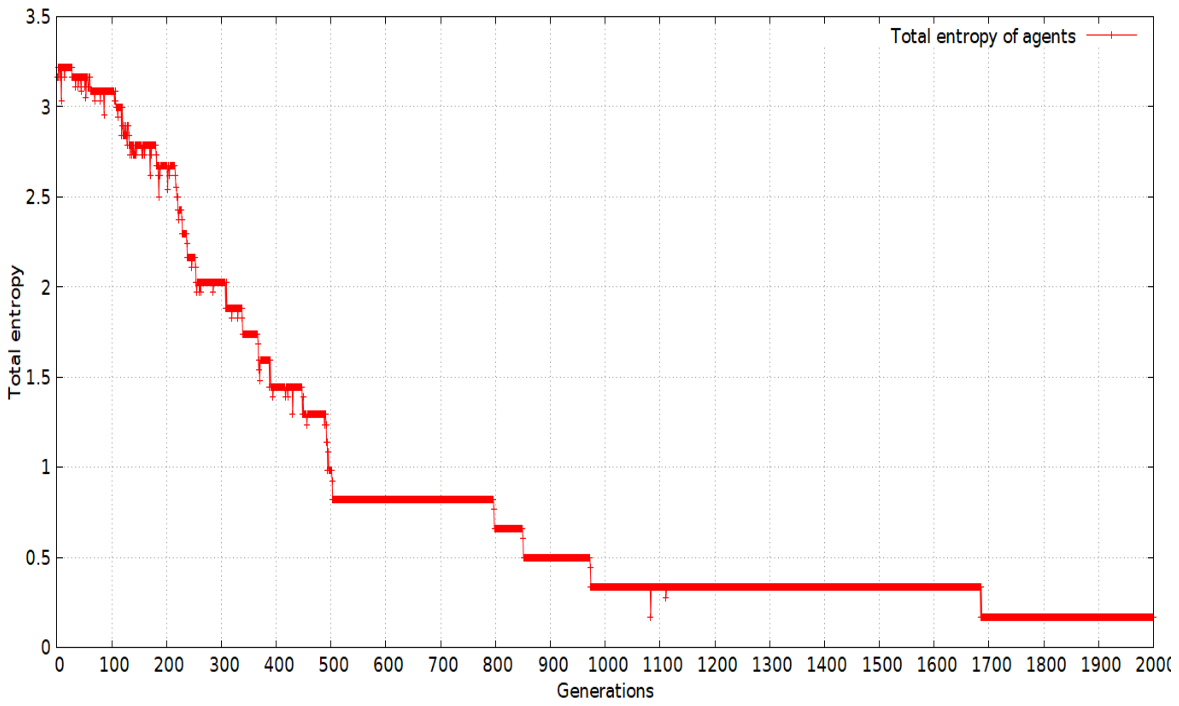}
\caption{Behaviour of entropy of solutions over generations)}
\label{fig_entropy_opt_inkscape}
\end{figure}

Fig. \ref{fig_entropy_opt_inkscape} shows the behaviour of the total information entropy of the solutions all the agents in overall system with generations. The total entropy is plotted along \emph{y}-axis and generations along \emph{x}-axis.
The behaviour of the graph may be approximated as truncated gaussian curve.
From the suboptimal solution in Fig. \ref{fig_total_resource_sub_opt_GA_inkscape}, the highest consumed resource (blue line) is roughly $1.2\times10^7$ at generation $2000$. If the same value $1.2\times10^7$ is taken along $y$-axis  from the optimality case (Fig. \ref{fig_total_resource_with_GA_inkscape}) which is roughly at generation 750 (along $x$-axis), the entropy at the same generation is 0.8 in Fig. \ref{fig_entropy_opt_inkscape}. Beyond generation 975 the entropy is almost constant and much lower as most agents achieve optimality. Thus, if the entropy is maintained within 0.5 and 0.8 the solutions are not substantially sub-optimal and at the same time more sustainable. This is probably the range where disruption and regulations can play a significant role in reducing resource consumption. Before generation 500 the solutions may be substantially sub-optimal.
\subsection{Utility of agents after achieving optimality}
\begin{figure}[ht]
\centering
\includegraphics[width=\columnwidth]{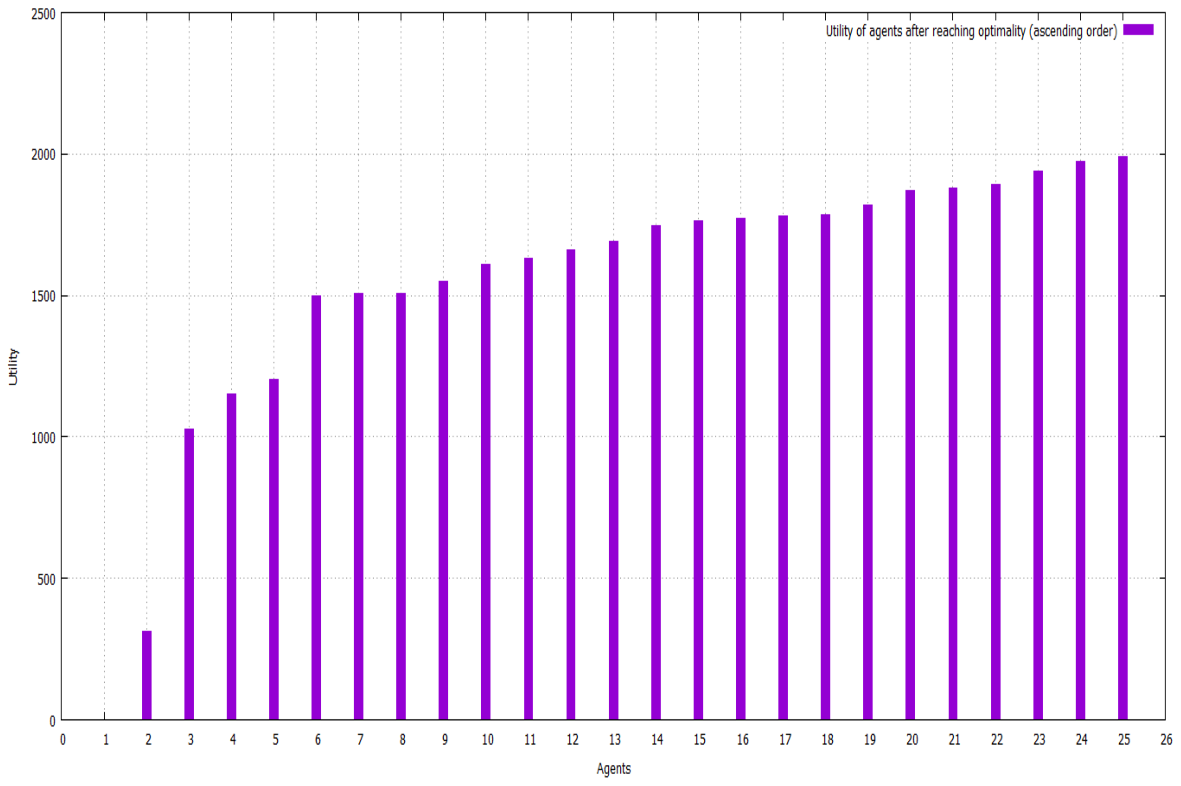}
\caption{Cummulative utility of agents after achieving optimality (ascending order)}
\label{fig_eco_divide_opt_inkscape}
\end{figure}

\begin{figure}[ht]
\centering
\includegraphics[width=\columnwidth]{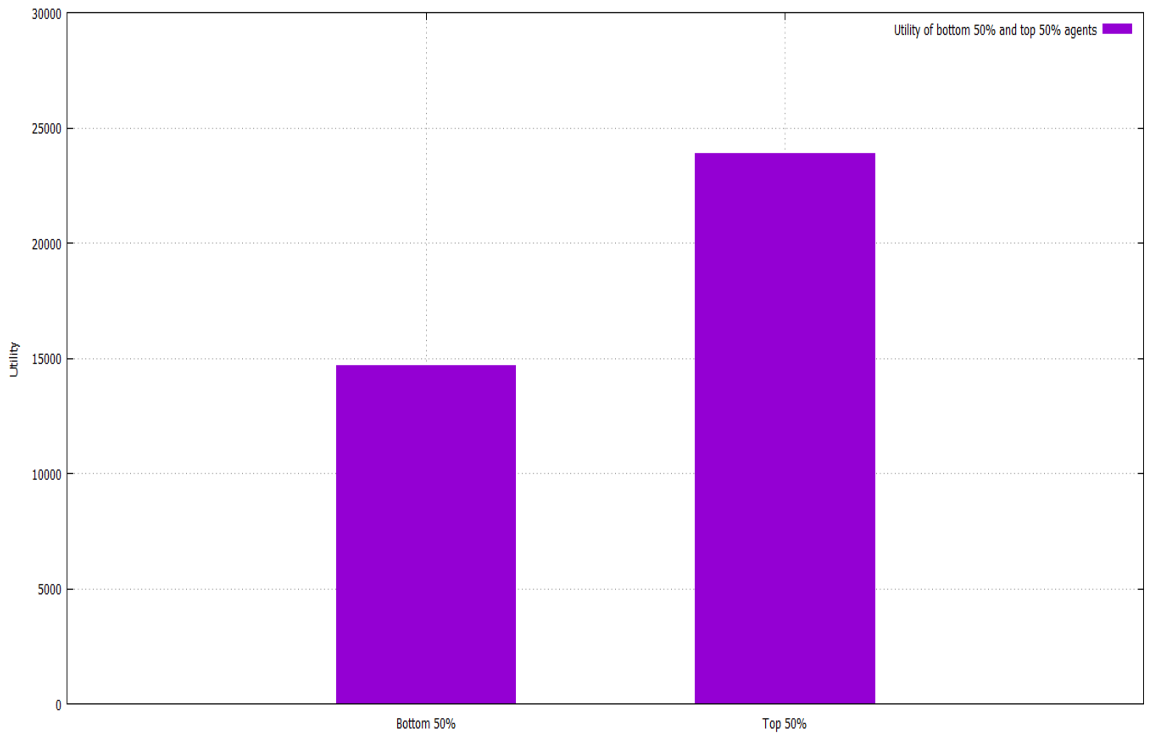}
\caption{Cummulative utility of bottom and top 50\% agents}
\label{fig_eco_divide_50_50_opt_inkscape}
\end{figure}

Agents who are early achievers of optimality tend to benefit over time (i.e., generations) assuming linear utility of using the global optimum. Fig. \ref{fig_eco_divide_opt_inkscape} shows the resources consumed (along \emph{y}-axis) by the agents (along \emph{x}-axis) in ascending order. It is evident that some of the agents gain much more than the others. If the cummulative resource consumptions of the top 50\% and the bottom 50\% agents are compared (Fig. \ref{fig_eco_divide_50_50_opt_inkscape}), it can be observed that difference is close to $40\%$. Thus, optimality could lead to economic divide among agents.

\begin{figure}[ht]
\centering
\includegraphics[width=\columnwidth]{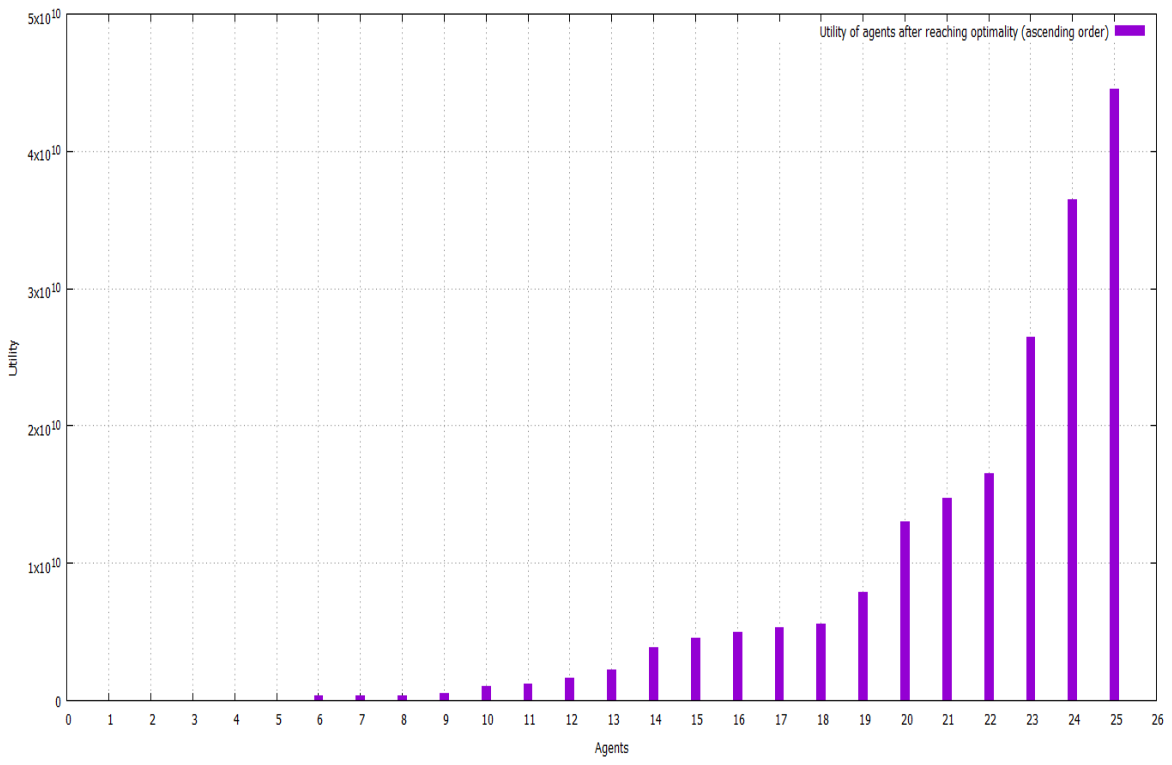}
\caption{Exponential utility - Cummulative utility of agents after achieving optimality (ascending order)}
\label{fig_monopoly_opt_inkscape}
\end{figure}

\begin{figure}[ht]
\centering
\includegraphics[width=\columnwidth]{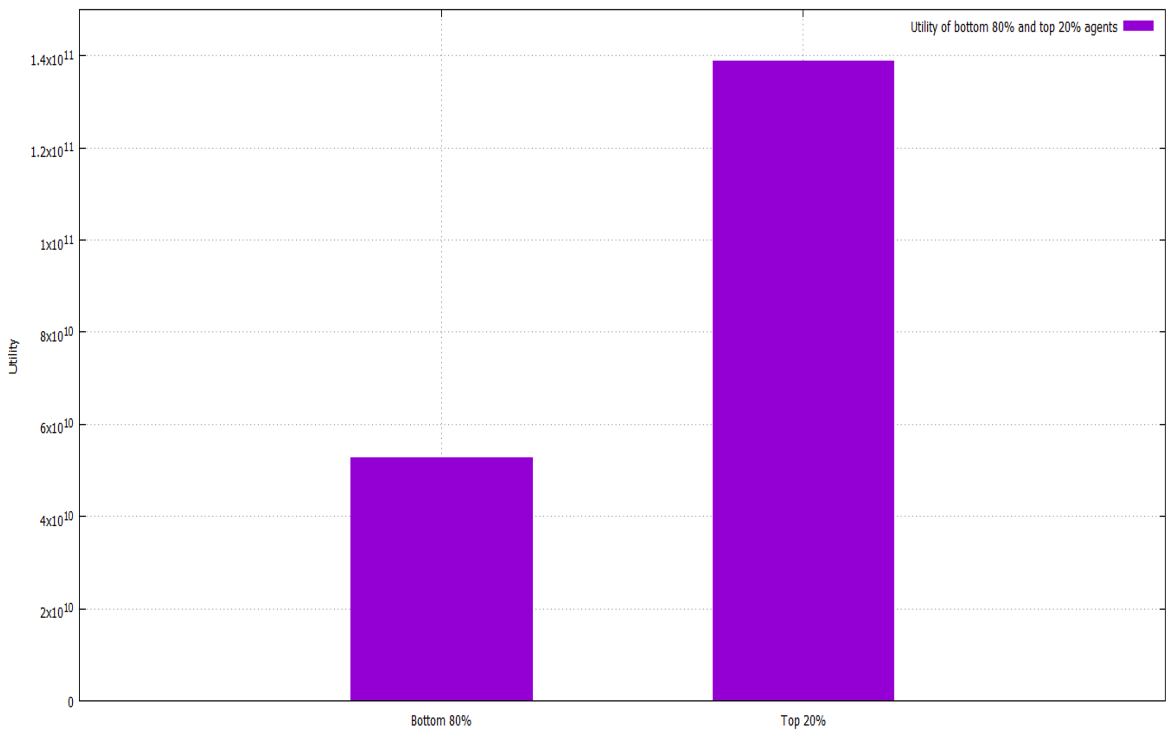}
\caption{Exponential utility - Cummulative utility of bottom 80\% and top 20\% agents}
\label{fig_monopoly_20_80_opt_inkscape}
\end{figure}

However, if the utility of using the global optimum is exponential with time then the gains of early achievers can be substantially higher (Fig. \ref{fig_monopoly_opt_inkscape}). The top $20\%$ agents' gains can actually be a whopping $162\%$ higher the remain 80\% agents. Thus, optimality can also lead to dominance $20\%$ of agents (Fig. \ref{fig_monopoly_20_80_opt_inkscape}).
\subsection{Utility of consumer and producer groups}
\begin{figure}[ht]
\centering
\includegraphics[width=\columnwidth]{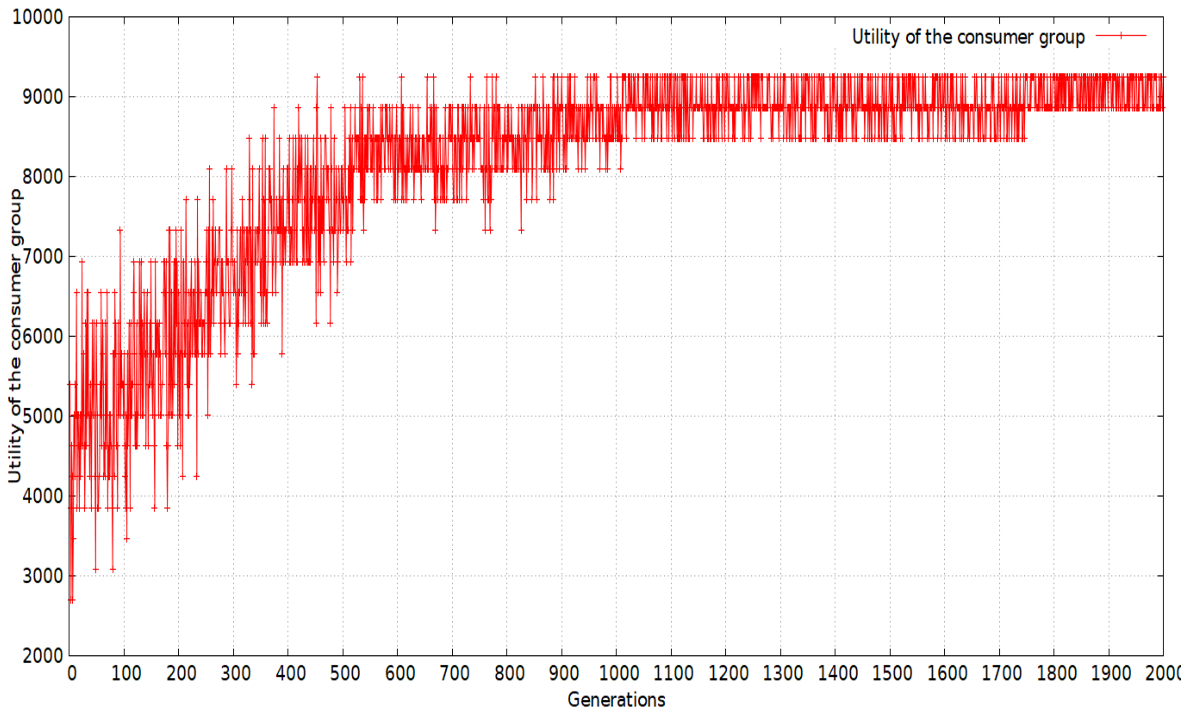}
\caption{Consumer Utility}
\label{fig_two_group_utility_top_total_opt_inkscape}
\end{figure}

Fig. \ref{fig_two_group_utility_top_total_opt_inkscape} shows the behaviour of the cummulative utility of the consumer group as they reach global optimality with time (generation). It can be observed that during the initial phase their is a increasing trend in the utility but saturates as more and more agents attain the global optimal solution similar to a monomolecular distribution. The variations in the total utility value also decrease with time.

\begin{figure}[ht]
\centering
\includegraphics[width=\columnwidth]{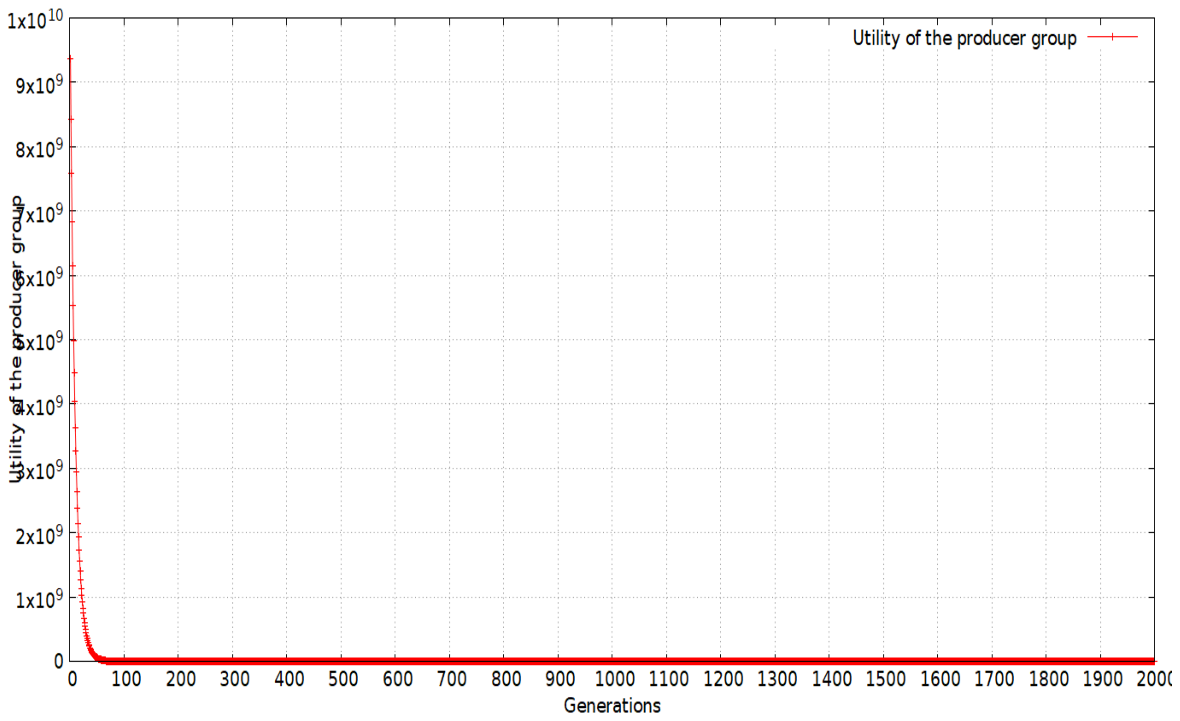}
\caption{Consumer Utility}
\label{fig_two_group_utility_bottom_total_opt_inkscape}
\end{figure}

The behaviour of the cummulative utility of the producer group is shown in Fig. \ref{fig_two_group_utility_bottom_total_opt_inkscape}. The figure shows a drastic decline in utility and stabilises quickly to a very low value depicting a power law. Thus, the consumer group seems to have a distinct gain out of optimality as time progresses.

\section{Possible solutions to accelerated tragedy of the commons}\label{section_way_possible_solutions}
Given that continuous and inevitable advancements in various fields mentioned earlier are ongoing, it becomes imperative to find strategies to address the situation of accelerated attainment of optimality (indicative of higher rationality) and the resulting swifter resource consumption.
One common approach to mitigate the tragedy of the commons is through the regulation of resource usages. This involves implementing rules and policies to manage and distribute resources more sustainably.
Another strategy is to focus on local solutions through basic research with the aim of efficiently utilizing local resources or identifying substitutes for scarce ones. This involves exploring innovative ways to adapt to resource limitations at the local level.
Disruptive innovation represents yet another avenue to introduce new techniques and technologies that can replace existing ones. Such innovations can lead to more efficient resource utilization and less environmental impact.
Interpreting the global optimal solution within the context of local conditions and customizing them to suit specific needs is crucial from a sustainability perspective. This involves tailoring global solutions to fit local contexts taking into account local constraints and requirements.
One of the major challenges for the future will be achieving a local optimal solution that is close to the global optimum while still maintaining diversity. Balancing the pursuit of efficiency with the preservation of diversity is essential for sustainable resource management.
In essence, addressing the consequences of accelerated optimization and resource consumption requires a multifaceted approach, including regulation, local adaptation, disruptive innovation and the careful balance of global and local solutions.

\section{Conclusion and future work}\label{section_conclusion}
The concept of rationality has long been a captivating subject of inquiry. The traditional notion of perfect rationality where agents are assumed to always make decisions that yield the global optimal solution has been challenged by the concept of bounded rationality. In bounded rationality, agents are recognized to have inherent limitations that prevent them from consistently reaching global optima. However, the ongoing advances in computing and various scientific and technological domains offer the potential to expand the boundaries of bounded rationality. This expansion may eventually lead to a state of perfect rationality where agents consistently achieve global optimum with the assistance of augmented machine intelligence applying computational rationality.
To investigate this idea, an agent-based computational model was employed in which agents endeavored to independently solve the 0/1-Knapsack problem using a trial-and-error approach. The results of this model revealed that agents when equipped with enhanced computing capabilities among other factors were able to reach the global optimum for the same problem more quickly. Paradoxically, this accelerated attainment of global optimum resulted in a faster depletion of shared resources, ultimately leading to what is known as the tragedy of the commons.
As agents achieve optimality, there is a drop in information entropy among the solutions of the agents. Further, clear economic divide emerges. Considering, two groups, one a producer and the other (the group agents searching for optimality) as consumer of the highest consumed resource, the consumers seem to gain more than the producers.
Indeed, the presence of diversity in solutions for the same problem, including local optima arising from bounded rationality appears to be a more sustainable approach when viewed from the perspective of computational sustainability. It offers resilience and adaptability in a dynamic environment.
However, it is crucial to acknowledge that there may be externalities or unintended consequences associated with agents using local sub-optimal solutions. These potential side effects should be carefully considered and studied in future research to develop comprehensive strategies.
Furthermore, the approaches suggested earlier for addressing the challenge of accelerated resource consumption, such as resource regulation, local solutions, disruptive innovation and customized global optimum hold significant promise. Investigating and implementing these strategies can help manage the impacts of emerging technical and scientific developments and promote a more sustainable and responsible use of resources.

\bibliographystyle{IEEEtran}
\bibliography{IEEEabrv,sample-base}

\end{document}